\documentclass[5p]{elsarticle}
\usepackage[english]{babel}
\usepackage{graphicx}
\usepackage{dcolumn}
\usepackage{bm}
\usepackage[latin1]{inputenc}
\usepackage{graphicx}
\usepackage{amsmath}
\usepackage{amssymb}
\usepackage{pifont}
\usepackage{epstopdf}
\usepackage{color}

\newcommand{\Da}{\textrm{Da}}

\begin{document}

\title{Theoretical modeling of catalytic self-propulsion}
\author{Tatiana V. Nizkaya}
\author{Evgeny S. Asmolov}
\author{Olga I. Vinogradova\corref{cor1}}
\ead{oivinograd@yahoo.com}

\address{Frumkin Institute of Physical Chemistry and
   Electrochemistry, Russian Academy of Sciences, 31 Leninsky Prospect,
   119071 Moscow, Russia}

\cortext[cor1]{}

\begin{abstract}
Self-propelling particles or microswimmers  have opened a new field of investigation with both fundamental and practical perspectives. They represent very convenient model objects for experimental studies of active matter and have many potential applications. Here, we summarize recent advances in theoretical description of  the self-propulsion of catalytic microswimmers that non-uniformly release ions, including its physical origins  and the current switch in focus to non-linear effects, geometric tuning and more. In particular, we show that the ionic self-propulsion always includes both diffusiophoretic and electrophoretic contributions,
and that non-linear effects are physical causes of a number of intriguing phenomena, such as the reverse in the direction of the particle motion in response to variations of the salt concentration or self-propulsion of electro-neutral particles.
 We finally suggest several remaining theoretical challenges in the field.
\end{abstract}

\maketitle

\section{Introduction}

In the last decade research on catalytic swimmers has rapidly advanced being strongly motivated by potential applications, such as drug delivery,  nano-robotics, and more. Such swimmers propel due to chemical ``fuel'' that generates non-uniformly distributed (over the surface) fluxes of the reaction products. The emerging local gradients of concentrations, in turn,  provide hydrodynamic stresses that cause a self-propulsion of the particle. Catalytic swimmers can take a multitude of diverse forms. However, so far mostly main geometries, such as sphere or cylinder have been favoured. Besides, the most studied examples remain the Janus particles that catalyze a chemical reaction and produce a uniform flux, but only at one (termed active) side of their surfaces.

Most catalytic swimmers represent charged particles immersed in an electrolyte solution. A solution containing ions builds up a so-called electrostatic diffuse layer (EDL) close to the charged particle, where the surface charge is balanced by the cloud of counterions. Such swimmers are
often referred to as ionic since they release ions. Ionic swimmers induce an electric field in the environment, which, in turn, affects their self-propulsion. Note that local ion concentration gradients and electric fields appear mostly within the EDL. Its extension is defined by the Debye length, $\lambda_D$, of a bulk solution, which is typically below a hundred of nm.

Despite the importance of ionic swimmers for various applications, fundamental understanding of several important aspects their self-propulsions did not begin to emerge until quite recently. Although there is growing theoretical literature in the area, many theoretical issues  still remain challenging.
Most attention has been focussed on the case of thin (compared to the particle characteristic dimension $a$) EDL that is well justified for microswimmers. In this case, it appears macroscopically that the liquid slips over the surface. The emergence of such an apparent slip flow can be used to calculate a self-propulsion velocity without tedious calculations. Another common assumption is that ion concentration and electric field
outside the EDL are only weakly perturbed by the particle, which allows one to derive linearized solutions ~\cite{moran2017phoretic}. Such solutions, however, do not describe the  situations of large electric potentials and ion fluxes.

The main thrust of recent research deals with the following
types of questions: (i) What is the physical origin of ionic self-propulsion and how does this depend upon the salt concentration in the bulk and electrostatic properties of particles? For example, is the self-propulsion of diffusiophoretic or electrophoretic nature, or both? Can the direction of particle motion be switched by adding salt and if so, what is the nature of such a reversal? (ii) How are transport properties such as diffusion of different ion species affect the velocity of the self-propulsion? (iii) What are the effects of neighbouring particles and confined geometry on
the self-propulsion? Since the answers to these
questions still cannot be obtained easily from experiment, it is not surprising that theoretical studies become especially important to provide fundamental input.

\begin{figure*}[h]
\centering
\includegraphics[width=1.3\columnwidth]{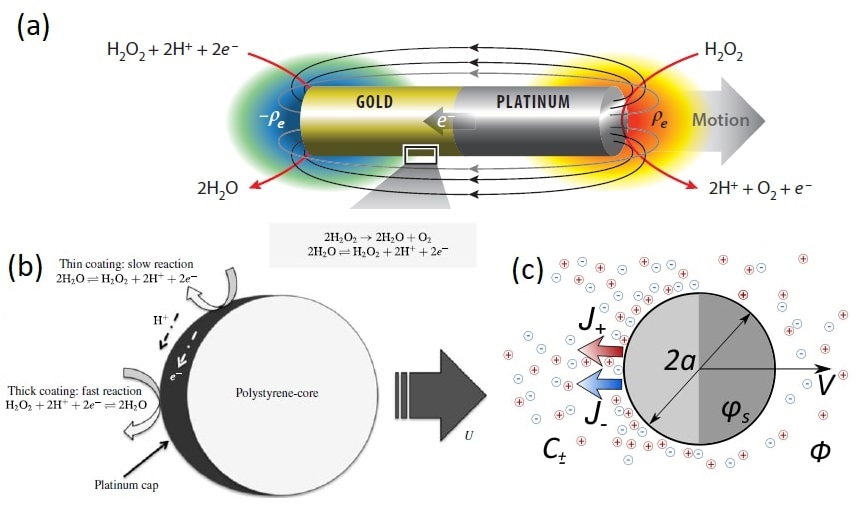}
\caption{Self-phoretic swimmers with ion release. (a) Bimetallic Pt-Au rod in hydrogen peroxide (reprinted from \cite{moran2017phoretic}). (b) Pt-PS Janus sphere in hydrogen peroxide (reprinted from \cite{ibrahim2017multiple}). (c) Janus sphere with equal fluxes of anions and cations (reprinted from \cite{asmolov2022self}).  }
\label{fig:types}
\end{figure*}

The present review attempts to give the flavour of some of the recent work in this
field. It is not intended to be comprehensive. Emphasis is placed on the main features of catalytic ionic swimmers of two main types and advances made during the last decade or so in the theory of their self-propulsion with the focus on the importance of non-linear phenomena.

\section{Models of swimmers and of emerging ion fluxes}\label{sec:types}

A phoretic mobility of the catalytic swimmers takes its origin in their ability to generate a fluid flow at the surface of the particle from local concentration gradients and/or electric fields along the surface.
Although the catalytic swimmers can be extremely complex, two main crude models are favoured~\cite{Wang2020practical,peng2022generic}:  particles that release solely one type of ions (typically, cations)  and those releasing both cations and anions (see Fig.~\ref{fig:types}).

In the former model swimmers release cations at one part of the surface and simultaneously absorb them at the other part. We refer them below to as swimmers of the first type (or Type I).
The classical examples include  bi-metallic (usually Au-Pt) rods and many analogous  systems that self-propel due to the catalytic decomposition of hydrogen peroxide~\cite{Paxton2004}. In this two-step reaction protons are released into the solution from
the Pt side and are absorbed by  the Au side. The electrons, in turns, are moving through the rod itself, by creating a closed electric circuit as shown in Fig.~\ref{fig:types}(a). The initially  neutral particle, thus, becomes negatively charged and the EDL is formed close to it. Induced in this layer electro- and
diffusio-osmotic flows then generate a propulsion of the rod in the direction of its Pt side.
Another example of swimmers of the first type is polystyrene spheres half-covered with platinum  (Pt-PS Janus particles)~\cite{Wang2020practical}, where protons are both released and adsorbed at the Pt side (at equator and at pole, respectively) as shown in Fig.~\ref{fig:types}(b)~\cite{ibrahim2017multiple}.

In the latter model it is postulated that the active part of swimmers
releases both anions and cations. Such swimmers will be referred to as those of second type (or Type II). The
 examples are photochemical motors propulsing by dissolution of $\textrm{AgCl}$~\cite{zhou2018photochemically}, biocatalytic motors that propel by enzyme-enhanced decomposition of organic compounds~\cite{hermanova2020biocatalytic} and more. Provided surface fluxes of anions and cations are equal, these particles do not produce an excess charge in the solution (see Fig.\ref{fig:types}(c)). Thanks  to unequal diffusivities of anions and cations, a charge imbalance can also arise in this case. As a result, an electric field can be induced both in the EDL and even at distances of the
order of particle size $a$. This second type of swimmers is much less theoretically understood, but has drawn much attention recently, mostly due to rapid development
of enzyme motors~\cite{hermanova2020biocatalytic,salinas2022recent}. Unlike bimetallic rods, enzyme
motors use bio-compatible ``fuel''~\cite{patino2018fundamental}, which  makes them quite attractive for various biomedical applications.

We stress that the swimmers of the first type are often termed  self-electrophoretic~\cite{moran2017phoretic,peng2022generic}, although it is by no means obvious that  a diffusio-phoretic contribution to a propulsion velocity can be ignored. By contrast, the swimmers of the second type are often referred to as self-diffusiophoretic~\cite{peng2022generic,zhou2018photochemically}, by ignoring that an electrophoretic contribution to their motion is neither negligible nor absent.
Already at this point it makes sense to argue that both types of swimmers should be termed self-phoretic since  electrophoretic and diffusio-phoretic contributions to their propulsion velocities must always be considered. We shall return to this issue later.

\begin{figure}[h]
\centering
\includegraphics[width=1\columnwidth]{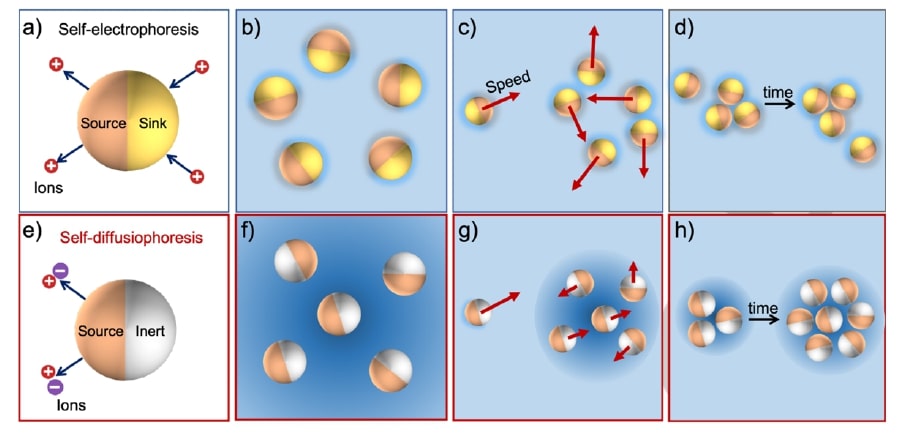}
\caption{Collective behavior of self-electrophoretic (a-d) and
self-diffusiophoretic (e-h) swimmers (reprinted from \protect\cite%
{peng2022generic}). }
\label{fig:collective}
\end{figure}

Note that understanding the type of particular swimmers is not always straightforward. A direct comparison of appropriate theory for a single particle and experiment is required, which is often not an easy task. However, very recently a simple alternative way to distinguish swimmers of different kinds has been proposed~\cite{peng2022generic}. The method is based on monitoring the collective behavior of swimmers (see Fig.~\ref{fig:collective}). Swimmers of the second type slow down in a crowded environment (Fig. \ref{fig:collective}(g)), unlike the ones of the first type (Fig.~\ref{fig:collective}(c)). They also form larger and more stable clusters (Fig.~\ref{fig:collective}(h)). The reason is that the particles of Type II act as sources of ions, so that the ensemble of such ion-releasing particles increases the ionic strength around them and thus reduces the phoretic mobility.

Finally, we mention that there are two approaches to modeling chemical activity  at the particle surface that produces the surface ion flux. Historically the first is the so-called kinetic-based model that was employed mostly for the swimmers of the first type~\cite{ibrahim2017multiple,moran2011electrokinetic,yariv2011electrokinetic,sabass2012nonlinear}. In this model the surface ion fluxes depend on the instant local concentrations. Thus, a detailed information on the rates of surface chemical reactions is required. In the flux-based models the surface ion fluxes are prescribed \cite{asmolov2022self,nourhani2015general,de2020self}. The advantage of  such a model for microswimmers is that their steady-state properties are amenable to much simpler, but still extremely detailed investigation.

\begin{table*}[h]
\centering
\begin{tabular}{lccr}
\hline
\textbf{References} & \textbf{Flux model} & \textbf{%
Method} & \textbf{Parameters} \\
\hline
\textbf{Type I} &  & & \\
\hline
  Moran et al. \cite{moran2011electrokinetic, moran2014role} &
kinetic-based & numerical & $\lambda<0.13, \text{Da}\sim0.5...5$ \\
Sabass et al.\cite{sabass2012nonlinear}  & kinetic-based & asymptotic
& $\lambda \ll1$, $\text{Da}=O(1)$ \\
Nourhani et al.\cite{nourhani2015general}& flux-based & asymptotic
& $\lambda \ll1$, $\text{Da}\ll1$ \\
Ibrahim et al.\cite{ibrahim2017multiple}  & kinetic-based &
asymptotic & $\lambda \ll1$, $\text{Da}\ll1$  \\ \hline
\textbf{Type II} &  & & \\
\hline
Zhou et al.\cite{zhou2018photochemically} & flux-based & numerical,
experiment & $%
\lambda \leq 1$, $\text{Da}=O(1)$ \\
de Corato et al.\cite{de2020self} & flux-based & numerical,
experiment & $\lambda \leq 1$, $\text{Da}\sim0.5...3000$ \\
Zhou et al.\cite{zhou2021ionic}  & flux-based & numerical,
experiment & $\lambda \leq 1$, $\text{Da}\sim0.3...3000$ \\
Asmolov et al.\cite{asmolov2022self} & flux-based & asymptotic & $%
\lambda \ll1$, $\text{Da}=O(1)$
\\ \hline
\end{tabular}%
\caption{Classification of the swimmer types, flux models, solution methods  and typical values of $\protect\lambda$
and  $\text{Da}$. }
\label{table}
\end{table*}

\section{Governing equations}\label{sec:GE}

One imagines a charged particle of an arbitrary shape  in contact with a reservoir of 1:1 electrolyte solution of permittivity $\epsilon$, a dynamic viscosity $\eta$, and concentration $c_{\infty}$. The Debye length of the bulk electrolyte solution is given by $\lambda _{D}=\left( 8\pi e^{2}c_{\infty }/\epsilon
k_{B}T\right) ^{-1/2}$, where $e$ is the
elementary positive charge, $k_{B}$ is the Boltzmann constant, and $T$ is the
temperature. The mobile ions  should adjust to one of the classical electrostatic boundary conditions at the particle surface. In the self-propulsion problems the condition of a constant surface potential $\phi_s$  is normally imposed. The latter is motivated  by analytical
convenience, and also by its widely accepted physical validity for many systems (e.g. metals).
Note that, here and below, the potentials are scaled by $k_{B}T/e$.  In the general case a particle can release both cations and anions, and
the surface fluxes of the released ions are not uniform, but not necessarily piecewise constant. The ions emerging at the surface diffuse into the bulk solution. A corollary is the concentration gradient arising at distances comparable to the particle size. Defining $D^{+}$ and $D^{-}$ as the diffusivity of released cations and anions, one can introduce a dimensionless parameter
\begin{equation}\label{eq:beta}
 \beta  = \dfrac{D^{+}-D^{-}}{D^{+}+D^{-}}.
 \end{equation}
If $\beta \neq 0$, in addition, an electric field is induced. This field slows down (speeds up) the diffusion of ions with greater (smaller)
diffusion coefficients.

The basic assumption of all analytic and numerical calculations is that for microparticles the Reynolds
number is always  small. Also, most of calculations considers that the Peclet number (Pe) is small for a typical diffusivity $D^{\pm }$ of the two ion species (i.e. cations and anions).
The latter assumption is well justified for most experimental systems reported so far, but could certainly become unrealistic for large and fast particles (such as for example  Au-Pt swimmers of $a \sim 100 \mu$m~\cite{moran2017phoretic}) or, say, when  surfactants of very low diffusivity are used as a ``fuel''~\cite{maas2016swimming}. However, previous theoretical investigations of the role of finite Pe have concerned the release of neutral solutes~\cite{michelin2014phoretic}, but  not of ions. By this reason,  here we discuss only the classical situation of Pe $\ll 1$. In this case the convective terms can be safely neglected and the Nernst-Poisson equations for the cation and anion fluxes reduce to~\cite{moran2017phoretic}
\begin{equation}
\mathbf{\nabla }\cdot \mathbf{J}^{\pm }=0,
\label{NP1}
\end{equation}%
\begin{equation}
\mathbf{J}^{\pm }=-D^{\pm
}\left( \mathbf{\nabla }c^{\pm }\pm c^{\pm }\mathbf{\nabla }\Phi \right) .
\label{NP2}
\end{equation}%
Here the
coordinates are scaled by a (characteristic) particle size $a,$  and $\Phi $ is the local electric potential. The first
term in Eq.(\ref{NP2}) represents the diffusion flux, while the second one is
associated with the flux due to ion electrophoresis.

An arising electric field satisfies the Poisson equation for the
electric potential,%
\begin{equation}
\nabla ^{2}\Phi =\frac{c^{+}-c^{-}}{2\lambda ^{2}c_{\infty }} . \label{pois1}
\end{equation}%
The new dimensionless parameter $\lambda =\lambda _{D}/a$ introduced above remains small for microparticles immersed in electrolyte solution since the nanometric Debye length  is small compared to their characteristic dimension.

Another important dimensionless parameter of the problem is the Dam\-k\"{o}hler number,
\begin{equation}
\mathrm{Da}=\frac{J a}{D c_{\infty }},
\end{equation}%
where $J$ is some characteristic (e.g. $J = \mathrm{max}\{ J^+, J^-\}$) flux of released from the surface ions
and $D$ is their diffusion coefficient. The Dam\-k\"{o}hler number is a characteristics of the excess
concentration of ions in the outer region relative to $c_{\infty }$, so that
\begin{equation}
c^{\pm }=c_{\infty }[1+O\left( \mathrm{Da}\right)].
\label{cc}
\end{equation}
We shall see at a later stage that even small $\mathrm{Da}$ can significantly impact a self-propulsion velocity.

\section{Methods of solution}\label{sec:NLM}

To find the migration (phoretic) velocity of the particle, quantitative calculations of the inhomogeneous concentrations and
potential are required. The system of the Nernst-Planck-Poisson equations \eqref{NP1}-\eqref{pois1} can be solved numerically or using asymptotic methods.
There is rather large literature describing a self-propulsion mechanism of these two types of ionic swimmers, and we mention in Table~\ref{table} what we believe are the more relevant contributions.

Before describing the results of calculations, it is
instructive to comment on information (on the flux models, investigation methods, and the range of parameters used) included in Table~\ref{table}. Let us first mention that in reality most theoretical papers has been addressed the particles of the first type. A large fraction of these deals with the modeling of chemical reactions at the surface, i.e. is using a kinetic-based approach. Swimmers of the second type begun to be studied only recently, and these investigations generally employed a flux-based model. We also note that while numerical calculations have been performed for a finite $\lambda$, all theoretical models of phoretic propulsion reported so far based on assumption of small $\lambda$. The Dam\-k\"{o}hler number, however, has been varied in a very large range. Theoretical work  mainly considered the case of small $\mathrm{Da}$, although in the experiments it is finite or even large. Most recent theoretical attempts, however, focussed on the situation of  finite $\mathrm{Da}$.

Theoretical work, thus, was always based upon an assumption of $\lambda \ll 1$. In this limit, an accurate solution to a
system of  equations for concentration and potential profiles can be found by using the method
of matched asymptotic expansions. In such an approach, the surrounding solution is conventionally divided into two regions with the dividing surface located at a distance $\lambda_D$ from the particle. The dividing surface has a potential $\Phi_s$ and concentration $c_s$.  The asymptotic expansions are then constructed in the inner and the outer regions of the length scales $\lambda _{D}$ and $a$, respectively. Essentially, the diffusio- and electro-osmotic flows are
generated in the very thin inner region, so that it macroscopically appears that the liquid slips over the surface with a velocity  $\mathbf{V}_{s}$.

By introducing a dividing surface and defining its potential $\Phi_s$ and  corresponding concentration of electrolyte $c_s$, the expression for the apparent slip velocity can be derived from the asymptotic solution for a flow in the
inner region~\cite{prieve1984motion,anderson1989colloid}
\begin{equation}
\mathbf{V}_{s}=\dfrac{\epsilon k_{B}^{2}T^{2}}{4\pi \eta e^{2}a}\left\{ \psi
\nabla_s \Phi _{s}+4\nabla_s(\ln c_{s}) \ln %
\left[ \cosh \left( \frac{\psi }{4}\right) \right] \right\} ,  \label{slip}
\end{equation}%
where $\nabla_s$ is the gradient operator along the particle surface and $\psi =\phi _{s}-\Phi _{s}$ is the potential drop in the inner region. The
first and the second terms in Eq.\eqref{slip} are associated with electro- and diffusio-osmotic
flows, respectively. Eq.\eqref{slip} includes the unknown yet $\Phi_s$ and $c_s$ that can be found from the outer solution.

The leading-order solution of (\ref{pois1}) for the (electroneutral) outer region is
$c^+=c^-\simeq c$.
The system of Eqs.\eqref{NP1} and \eqref{NP2} can then be simplified to give \cite{ibrahim2017multiple,nourhani2015general}
\begin{equation}
\Delta c=0,  \label{dif}
\end{equation}%
\begin{equation}
\nabla \cdot \left( c\nabla \Phi \right) =0,  \label{pot2}
\end{equation}%
with boundary conditions at infinity $c \left( \infty
\right) =c_{\infty}$, $\Phi \left( \infty
\right) =0$, and also with conditions for the ion fluxes at the dividing surface. We stress that
although the outer region is
electroneutral to the leading order in $\lambda $, a finite $\Phi$ is induced provided $\mathrm{Da}=O(1)$. Consequently, $\Phi _{s}$ is finite and affects $\psi$.

The methods of solution of the linear Laplace equation (\ref{dif}) are
well-known. For a single spherical particle the concentration is expressed
in terms of Legendre polynomials. The concentration disturbances decay as $r^{-2}$ for swimmers of the first type where a net ion flux is zero, and as $r^{-1}$ for those of the second type. However, the solution of a non-linear equation for the electric
potential \eqref{pot2} still remains a challenge.

The emergence of an apparent slip flow provides hydrodynamic stresses that cause the self-propulsion of the particle at some direction $x$.
Note that the swimmer phoretic velocity $v_p$ can be found by
using the reciprocal theorem~\cite{teubner1982,masoud2019}, i.e. it is not necessary to
solve the classical Stokes equations. In this approach the swimmer velocity is given by
\begin{equation}
v_{p}=-\frac{1}{6\pi }\int \mathbf{f}\cdot \left( \mathbf{v}_{1}-%
\mathbf{e}_{x}\right) dV,  \label{v_rec}
\end{equation}%
where $\mathbf{f}$ is an electric body force, $\mathbf{e}_{x}$ is a unit
vector in the swimming direction, and $\mathbf{v}_{1}$ represents the velocity field for
the same particle that translates with the velocity $\mathbf{e}_{x}$ in a
stagnant fluid (without volume forces). Note that
the integral is evaluated over the whole
fluid volume.  Eq.(\ref{v_rec}) is valid
for any $\lambda,$ but if $\lambda \ll 1,$ the particle velocity may be determined from~\cite{prieve1984motion,anderson1989colloid}
\begin{equation}
v_{p} \simeq -\frac{1}{S}\int\left( \mathbf{V}_{s}\cdot \mathbf{e}%
_{x}\right) dS . \label{vpx}
\end{equation}%
This surface integral over particle surface is associated with the
contribution of the inner region only~\cite{prieve1984motion,anderson1989colloid}. It has recently been shown that Eq.~(\ref{vpx}) should also include an additional, associated with the outer region, term
$\frac{1}{6\pi }\int\Delta \Phi \mathbf{\nabla }\Phi
\cdot \left( \mathbf{v}_{1}-\mathbf{e}_{x}\right) dV$~\cite{asmolov2022self}.  The latter, however,
is quite small for the swimmers of the second type. It would be of interest to investigate its contribution in the case of swimmers of the first type too.

Prior work mostly addressed the case of an axisymmetric distribution of slip velocity that results in the alignment of the propulsion direction $\mathbf{e}_x$  with the axis of symmetry of the swimmer. When the axial symmetry is broken, the particle gains a rotational velocity~\cite{anderson1989colloid},
\begin{equation}
\mathbf{\omega}_p \simeq\dfrac{3}{2a}\frac{1}{S}\int\left( \mathbf{V}_{s}\times \mathbf{n}\right) dS,  \label{v_s}
\end{equation}%
where $\mathbf{n}$ is a unit vector normal to particle surface. Such a rotation leads to helical trajectories of self-propelling particles~\cite{ebbens2018catalytic}.  

%

\subsection{Finite Da (nonlinear problem)}\label{sec:NLM}

\begin{figure}[h]
\centering
\includegraphics[width=1.\columnwidth]{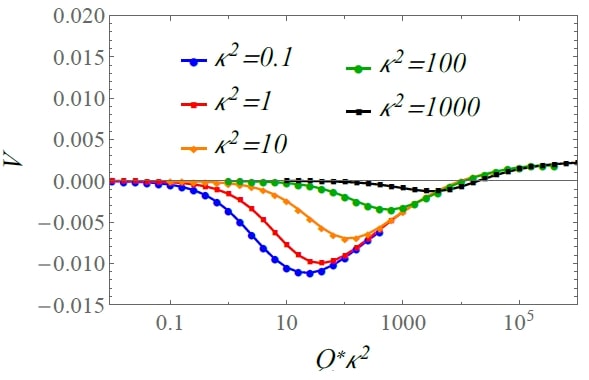}
\caption{Propulsion velocity vs. $Q^*\kappa^2=\text{Da}\lambda^{-2}$ computed for a positively charged particle using  several $\kappa=\lambda^{-1}$ and $\beta =-1/3$. (reprinted from \protect\cite{de2020self}).}
\label{nonlinear}
\end{figure}

The case of finite $\text{Da}$ is usually solved numerically, even for small $\lambda$. Figure~\ref{nonlinear} shows the representative results of such calculations~\cite{de2020self}. In this work the Nernst-Planck-Poisson equations have been solved numerically for a wide range of $\text{Da}$ with the consequent calculation of the self-propulsion velocity using the reciprocal theorem. The propulsion velocity is plotted as a function of $\text{Da} \lambda^{-2}$. It can be seen that the swimming speed  depends nonlinearly on $\text{Da} \lambda^{-2}$. In particular, at large fluxes, the reversal in the swimming direction is predicted. This phenomenon has been later confirmed experimentally~\cite{zhou2021ionic}.
At very large $\text{Da} \lambda^{-2}$ the numerical curves in Fig.~\ref{nonlinear} obtained for different $\kappa = \lambda^{-1}$ converge into a single one. The theoretical understanding of this result has not yet been proposed and awaits for more data.

\begin{figure}[h]
\centering
\includegraphics[width=1.\columnwidth]{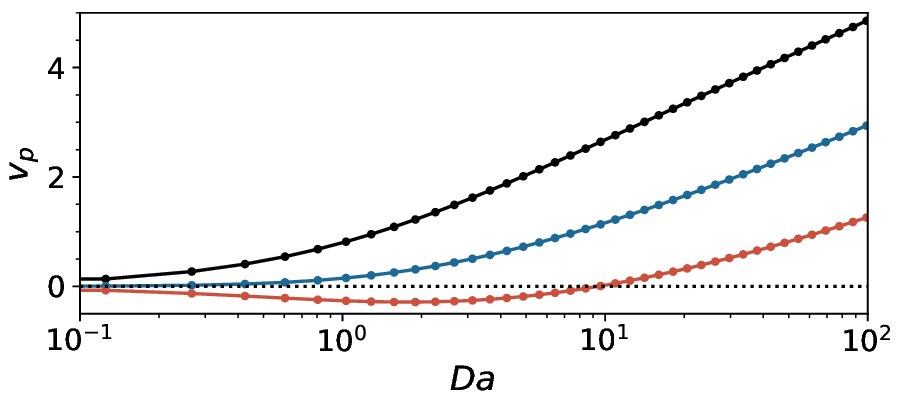}
\caption{Particle velocity vs. $\Da$ calculated using $\phi_s=2$, $0$, $-2$ (from top to bottom) and $\beta=0.5$ (plotted using data from \protect\cite{asmolov2022self}).}
\label{nonlinear2}
\end{figure}

Recently Eq.\eqref{pot2} has been solved analytically for  swimmers of an arbitrary shape. It has been found that the outer potential for particles of the first type
is given by~\cite{asmolov2022MDPI}%
\begin{equation}
\Phi =\ln (c/c_{\infty }),  \label{F1}
\end{equation}%
which is equivalent to the Boltzmann distribution for
anions $c=c_{\infty }\exp (\Phi )$.
For the particles of the second type the solution reads~\cite{asmolov2022self,asmolov2022MDPI}%
\begin{equation}
\Phi =-\beta\ln (c/c_{\infty }),
\label{fs2}
\end{equation}%
that is equivalent to the condition of $\mathbf{J}^+=\mathbf{J}^-$ over the whole domain.

Eqs.\eqref{F1} and \eqref{fs2} allow us to calculate the
gradients of electric field and concentration in Eq.\eqref{slip} and to conclude that they are of the same orders. Thus, at finite $\psi$
the electro- and diffusio-osmotic contributions to the slip velocity for particles of both types are comparable. However, if the potential drop in the inner region
is small,  the first term in Eq. \eqref{slip} grows linearly with $\psi$, while the second one is
quadratic in $\psi$. Therefore, in this limiting case the electro-osmotic contribution to $\mathbf{V}_{s}$ dominates for both types of swimmers, so that we deal with a self-electrophoretic motion of the particle.

Figure~\ref{nonlinear2} shows the results of calculations of the velocity $v_p$ of a spherical Janus swimmer of the second type from Eq.\eqref{vpx}. The calculations are made using data of \cite{asmolov2022self}. The apparent slip velocity $\mathbf{V}_{s}$ has been found from Eq.\eqref{slip} using $\Phi$ given by \eqref{fs2}.
It can be seen that the propulsion velocities vary  with \text{Da} nonlinearly. We remark that the swimming speed becomes finite even for uncharged particle. Another important result is that in the case of positive $\phi_s$, the swimming direction reverses on increasing $\text{Da}$. Finally, we note that at large $\text{Da}$  in this semi-log scale the growth of $v_p$ with Da appears as linear indicating that the propulsion velocity scales with $\ln \mathrm{Da}$. An approximate solution for $v_p$ in this limit has not yet been found, but one can suggest that such an asymptotics is likely caused by a high  $\Phi_s \propto \beta \ln \mathrm{Da}$ at the dividing surface~\cite{asmolov2022self}. One can then speculate that $\Phi_s$  gives a major contribution to a potential drop $\psi$ that determines $\mathbf{V_s}$ (see Eq.\eqref{slip}).

\subsection{Small Da (linear problem)}\label{sec:LM}

Since $c=c_{\infty}[1+O\left( \mathrm{Da}\right) ]$,  the excess concentration is small when
$\mathrm{Da}\ll 1$. In this situation Eq.\eqref{pot2} can be linearized,
and the problem is reduced to the solution of the Laplace equation. The outer potential for the particle of type I is then given by~\cite{ibrahim2017multiple}%
\begin{equation}
\Phi =c/c_{\infty
}-1=O\left( \mathrm{Da}\right) .  \label{fs3}
\end{equation}%
Therefore, $\psi \simeq \phi _{s},$ and the slip velocity given by Eq. \eqref{slip} is proportional to $\mathrm{Da}
$ and to the local concentration gradient.
The apparent slip velocity can thus be written as~\cite{golestanian2007designing,popescu2016self}
\begin{equation}
\mathbf{V}_{s} \simeq -\mu \nabla_s c_s,  \label{v_s}
\end{equation}%
where $\mu$ is the so-called phoretic mobility, an intrinsic property of the surface, which does not depend on $c_s$.
There is a large literature describing attempts to provide a satisfactory theory of self-propulsion of an isolated swimmer by employing approximate Eq.\eqref{v_s}. We consider
these to be important contributions, but note that the most intriguing phenomena, such as a reversal of the swimming direction or a self-propulsion of neutral particles cannot be predicted within a linear model. However, linear theory has been already applied for a number of important problems, which still await for a nonlinear treatment. We mention below what we believe are the more relevant contributions.

One of the hot subjects includes the optimization of the swimming speed that, of course, depends on the particle shape and distribution of chemically active sites~\cite{Guo2021optimal,Ibrahim2018shape,yariv2020self}. By applying the minimum dissipation theorem, the bounds on  viscous dissipation can be calculated~\cite{Nasouri2021minimum}, and by  analyzing them, one can optimize the swimmer geometry.

\begin{figure}[h]
\centering
\includegraphics[width=1.\columnwidth]{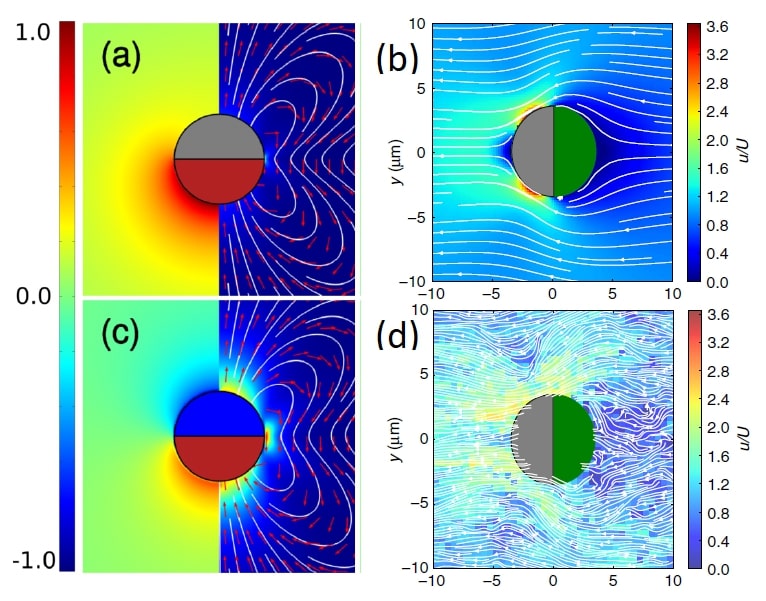}
\caption{(a,c) Concentration (left) and velocity (right) fields around the swimmers of Types II  (a) and I (c) %
(reprinted from \protect\cite{kreissl2016efficiency}).  (b,d) Velocity field around a Pt-PS Janus
sphere, calculated numerically (b) and measured in the experiment (d) (reprinted from \cite{campbell2019experimental}). }
\label{fig:velfield}
\end{figure}

Another area of active research is the collective behavior of swimmers and their motion in confined geometry~\cite{popescu2018effective1,kanso2019phoretic,singh2019,traverso2020hydrochemical,nasouri2020,liebchen2021interactions}. Note that to control the confined swimmer motion one can also use chemically patterned walls~\cite{uspal2019active,chen2018chemically}. The interaction of particles arises, thanks to disturbances of concentration and velocity fields (induced by neighboring particles or by walls). The calculation of these fields
(by solving the Laplace~\eqref{dif} and Stokes equations~\cite{happel2012low,popescu2018effective2,rojas2021hydrochemical})  is certainly one of the main thrust of recent research.
 Figure~\ref{fig:velfield}(a,c) illustrates the concentration distribution (left) and the fluid velocity field (right) near  the typical Janus swimmers, computed within the linear model~\cite{kreissl2016efficiency}. The results for a swimmer of Type I are shown in Fig.~\ref{fig:velfield}(c). The regions of concentration enrichment and depletion are well pronounced  since all released ions are absorbed by the opposite part of the particle (the total ion flux in a steady regime vanishes). For swimmers of Type II (Fig.~\ref{fig:velfield}(a)) the total flux is positive and particles act as constant sources of ions. As a result, concentration disturbances are positive and extend to larger distances. For both types of swimmers the velocity field looks quite similar at distances of the order of $a$, but in the vicinity of the particle the velocity becomes very large only for swimmers of the first type. Similar numerical calculations have been made for Pt-PS Janus swimmers, together with experimental verification~\cite{campbell2019experimental}. The numerical and experimental velocity fields are shown in Fig.~\ref{fig:velfield}(b) and (d). We see that they are in a good agreement. All these features of the velocity field lead to different collective behavior of swimmers~\cite%
{peng2022generic,liebchen2021interactions} (see also Fig.~\ref{fig:collective}).

\section{Conclusion}

As emphasized in the introduction, this article has concentrated on the self-propelling properties of  ionic catalytic swimmers. Theory has made striking advances during the last years leading to
predictions and interpretation of novel nonlinear phenomena, such as the reversal of the swimming direction and more.
A fundamental understanding of what type of phoretic motion (i.e. electro- and/or diffusiophoresis)
might be expected to occur in different situations has been established.
A significant progress has also been made in optimizing the swimmer shapes
and in understanding of the complex collective behavior, although within a linear theory only.  The time is probably right for detailed investigations within a nonlinear theory of more complex and realistic systems, such as a collective self-propulsion or swimmers in confined geometry.

Another fruitful direction could be the nonlinear analysis to guide the optimization of the particle shape, surface fluxes and other parameters in order to provide required swimming speed and direction.

\section*{Acknowledgements}
This work was supported by the Ministry of Science and Higher Education of the Russian Federation.


\end{document}